\documentclass[
reprint,
amsmath,
amssymb,
aps,
prapplied,
superscriptaddress
]{revtex4-2}

\usepackage{graphicx} 
\usepackage{dcolumn}  
\usepackage{bm}       
\usepackage[version=4]{mhchem}
\usepackage[utf8]{inputenc}
\usepackage[T1]{fontenc}
\usepackage{siunitx}  
\usepackage{upgreek}

\makeatletter

\newcommand{\largeSection}[1]{%
    \begingroup%
    \LARGE%
    \section*{#1}%
    \endgroup%
}

\newcommand{\suppnumbering}{%
  \setcounter{section}{0}
  \setcounter{subsection}{0}
  \setcounter{figure}{0}
  \setcounter{table}{0}
  \setcounter{equation}{0}
  \renewcommand{\thesection}{S\Roman{section}}
  \renewcommand{\thesubsection}{\thesection.\Alph{subsection}}
  \renewcommand{\thefigure}{S\arabic{figure}}
  \renewcommand{\thetable}{S\arabic{table}}
  \renewcommand{\theequation}{S\arabic{equation}}
}

\makeatother

\begin{document}

\preprint{APS/123-QED}

\title{Spin-Wave Phase Shifter Controlled by a Domain Wall Racetrack}

\author{Uladzislau Makartsou}
\email{ulamak@amu.edu.pl}
\affiliation{Institute of Spintronics and Quantum Information, Faculty of Physics and Astronomy, Adam Mickiewicz University, Uniwersytetu Poznańskiego 2, 61-614 Poznań, Poland}
\author{Olena Tartakivska}
\affiliation{Institute of Spintronics and Quantum Information, Faculty of Physics and Astronomy, Adam Mickiewicz University, Uniwersytetu Poznańskiego 2, 61-614 Poznań, Poland}
\author{Paweł Gruszecki}
\affiliation{Institute of Spintronics and Quantum Information, Faculty of Physics and Astronomy, Adam Mickiewicz University, Uniwersytetu Poznańskiego 2, 61-614 Poznań, Poland}
\author{Anton Lutsenko}
\affiliation{NanoSpin, Department of Applied Physics, Aalto University School of Science, PO Box 15100, FI-00076 Aalto, Finland}

\author{Sebastiaan van Dijken}
\affiliation{NanoSpin, Department of Applied Physics, Aalto University School of Science, PO Box 15100, FI-00076 Aalto, Finland}
\author{Volodymyr V. Kruglyak}
\affiliation{University of Exeter, Stocker Road, Exeter, EX4 4QL, United Kingdom}
\author{Maciej Krawczyk}
\affiliation{Institute of Spintronics and Quantum Information, Faculty of Physics and Astronomy, Adam Mickiewicz University, Uniwersytetu Poznańskiego 2, 61-614 Poznań, Poland}

\date{\today}

\begin{abstract}  
We propose a spin-wave phase shifter controlled using a domain-wall racetrack. The concept is demonstrated using micromagnetic simulations of a Permalloy domain-wall racetrack placed above a YIG film. The stray field from pinned domain walls modifies the internal magnetic field in the YIG region under the racetrack. This leads to a local change of the spin-wave wavelength and thereby enables control of the phase accumulated by Damon–Eshbach spin waves propagating through the region. 
 Moving domain walls on the racetrack, the same physical structure can provide phase shifts of up to $\pm90^\circ$, 
without changing the waveguide geometry. A model based on the semiclassical approximation confirms that the phase shift is dominated by the domain-wall-induced stray field.
These results suggest a route toward a compact programmable spin-wave phase shifter for interference-based magnonic circuits for information processing. Moreover, the demonstrated magnonic device integration with a magnetic domain-wall racetrack can lead to its application in in-memory computing.
\end{abstract}
\maketitle

\section{Introduction\label{Sec:Introduction}}
Spin waves (SWs) are wave-like collective excitations of magnetically ordered systems. The use of SWs and their quanta, magnons, for information processing is a major practical target for research in magnonics \cite{Kruglyak_2010}. Information can be encoded either in the amplitude or phase of propagating SWs, whose interference represents a key step in the processing of this information \cite{barman2021magnonics}. In this respect, efficient control of the SW phase is especially important. The relative phase of the input SW signals determines whether the output signal is enhanced or suppressed, before being mapped onto logic states \cite{Hertel2004, Schneider2008, Lee2008, Kanazawa2017}. Controlling the input phase is also a prerequisite for the operation of magnonic lookup tables (LUTs) based on the self-imaging (Talbot) effect in a magnonic diffraction grating, where phase-encoded input information is converted into amplitude contrast at the output \cite{Golebiewski2022,10.1063/5.0195099,PhysRevApplied.22.014038}. 
SW interference and phase control play a key role in the emerging devices for unconventional magnonic computing. The leading concepts of nanoscale magnonic neural networks use nonlinear SW interference for pattern-recognition and decision-making tasks, promising low-power scalable hardware architectures \cite{Papp2021, Wang2021InverseDesign}. Thus, development of compact efficient SW phase shifters is essential for both conventional (i.e. based on binary logic and arithmetic) and emerging unconventional (e.g. wave-based, neuromorphic, reservoir, non-Boolean, etc) magnonic computing. More generally, spatially varying phase shifts can be used to reshape SW wavefronts and thereby control the propagation direction. For example, SW focusing was demonstrated by modifying the exchange coupling at an interface between two ferromagnetic films \cite{Zelent2019}, while anomalous refraction and signal guiding in curved multimode magnonic waveguides were achieved using a spatial variation of the saturation
magnetization \cite{mieszczak2020anomalous}.

Proposed approaches for SW phase manipulation can be broadly divided into two categories: resonant and non-resonant. In the former, the SW phase is controlled via coupling to a local resonant element or mode \cite{Kruglyak2021}. Its leading experimental realization employs modes circulating in a magnonic Fabry--P\'{e}rot resonator \cite{Qin2021_nature,Lutsenko2025} composed of a ferromagnetic metallic stripe fabricated on top and dipolar-coupled to an yttrium-iron garnet (YIG) film. Coupling of propagating SWs to one such mode yields a \(\pi\) phase shift (at about 65\% of amplitude transmission), programmable by switching the orientation of the stripe's magnetization~\cite{Qin2021_nature, Lutsenko2025}. Numerical demonstrations of programmable SW-phase shifters include the use of the stripe's Kittel~\cite{Au2012} and dark~\cite{Fripp2021b} modes, their control by spin currents~\cite{Zhang_2020}, construction of Fabry--P\'{e}rot resonators using stripes with a magnetization lower than of the medium \cite{smigaj2023modal}, and the use of the reflection geometry in magnonic Gires--Tournois interferometers~\cite{sobucki2021}. 
The resonant schemes above use ferromagnetic stripes with widths smaller than the incident SW wavelength, which enables an effective device miniaturization.

Among non-resonant approaches is the numerical demonstration of a phase shift acquired by SWs when passing through a magnetic domain wall~\cite{Hertel2004}. In a stripe-based one-dimensional magnonic crystal, the phase of the transmitted SW can be modified by reversing the magnetization of a single ferromagnetic stripe \cite{Baumgaertl2021}. More generally, the SW phase can be tuned by a graded magnonic index, i.e. a continuous variation of the internal magnetic field \cite{Kostylev2007}, geometry~\cite{Dobrovolskiy2019}, or any parameter that determines the local SW dispersion and so the SW wavelength. Of special importance for energy efficiency of the technology is the ability to control the SW phase by an electric field / voltage \cite{Qin2021,Liu2012,Petrillo2024} and by non-volatile, yet reversible means \cite{Au2012,Baumgaertl2021,Cocconcelli2025}.


\begin{figure}[ht]
  \centering
  \includegraphics[width=0.93\columnwidth]{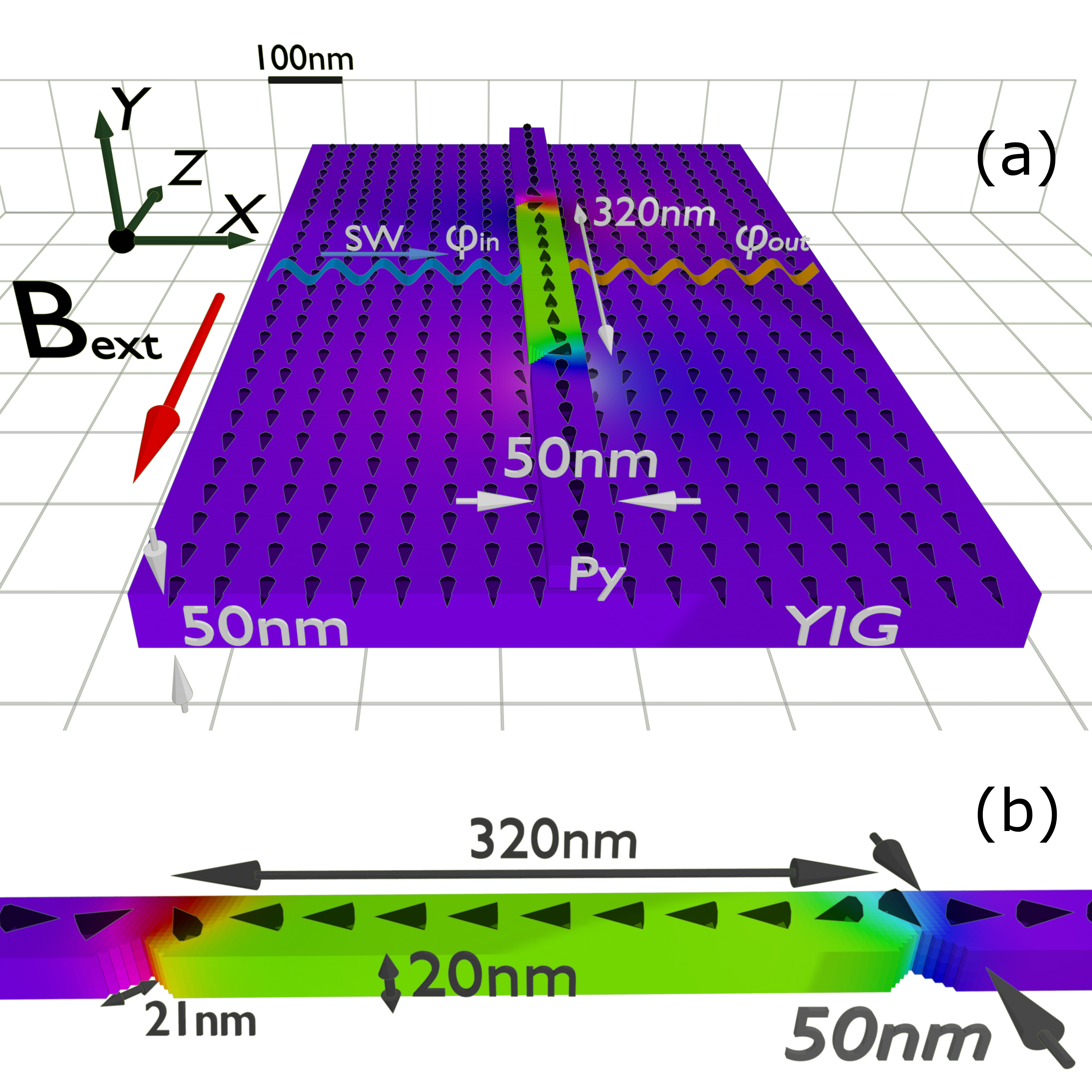}
\caption{Geometry and magnetic configuration of the simulated phase shifter. (a) Three-dimensional view of the model: A Py racetrack of \SI{50}{nm} width and \SI{20}{nm} thickness is separated by \SI{5}{nm} from a \SI{50}{nm} thick YIG film. A static bias magnetic field of \(B_{\mathrm{ext}}=\SI{-15}{mT}\) is applied along the \(z\)-axis. SWs propagate in the YIG film along the \(x\)-direction. The SW phase evaluated before and after the Py stripe is shown by \(\varphi_{\mathrm{in}}\) and \(\varphi_{\mathrm{out}}\), respectively. Black arrows indicate the equilibrium magnetization in YIG. (b) Micromagnetic state of the Py stripe: The magnetization in the central domain between the two notches located \( \SI{320}{nm}\) apart is antiparallel to that in the outer parts of the stripe.}
\label{Fig:structure}
\end{figure}

Magnetic domain-wall racetrack structures operate through the current-controlled displacement of magnetic domain walls (DWs) between pinning sites created by notches in a soft magnetic stripe ~\cite{Parkin2008,Bahri2019}. Each DW represents a compact source of a strong magnetic field, which lends itself readily for a local modification of the dispersion and phase of SWs in a nearby magnonic medium. Not only does this suggest a route toward compact SW phase shifters, but also it yields us an opportunity to integrate the non-volatile memory function of magnetic DW racetracks with the emerging unconventional computing architectures of magnonics. The high linear density of DWs hosted in a single racetrack is appealing for increasing both the density of magnonic circuits and the dimensionality and complexity of interference-based  neuromorphic magnonic architectures \cite{Papp2021, Wang2021InverseDesign, Kruglyak2021}.


In this work, we propose and numerically model SW phase-shifters formed by integration of a SW waveguide and a magnetic DW racetrack. The racetrack represents a Permalloy (Py) stripe placed above a YIG film or waveguide magnetized parallel to the stripe's axis (Fig.~\ref{Fig:structure}). Positions of DWs in the racetrack are fixed by two notches. A closely related DW-position-controlled SW phase modulator based on dipolar coupling in a Bi waveguide system was recently proposed~\cite{Mortada2026_DW}. Our approach embeds a phase-control mechanism into an established current-driven racetrack-memory architecture with notched DW pinning sites~\cite{Parkin2008,Bahri2019}.
In this geometry, the two neighboring DWs have opposite magnetic charges. The charges produce a stray magnetic field that locally modifies the internal field and thus the wavelength of SWs propagating in YIG underneath the magnetic domain between the two DWs. As a result, the SWs accumulate an additional phase, and the structure acts as a SW phase shifter. 
The two possible domain configurations in the racetrack with respect to the magnetization in YIG lead to opposite phase shifts of about \(90^\circ\), with a smooth frequency dependence over a broad frequency range, while keeping the transmitted amplitude close to the reference level of the YIG film without the racetrack. The results of the micromagnetic simulations agree with those obtained using a model of the device developed in the semi-classical (WKB) approximation. 

\section{Geometry and the simulation method\label{Sec:SimGeometryParams}}

The geometry and magnetic configuration of the considered system are shown in Fig.~\ref{Fig:structure}. A Py stripe is separated by 5 nm from a \SI{50}{nm} thick YIG film. The stripe has a width of \SI{50}{nm} and a thickness of \SI{20}{nm}. Two notches of triangular shape and \SI{21}{nm} side length are introduced into the stripe to stabilize two DWs \SI{320}{nm} from each other.  A static external magnetic field of \(B_{\mathrm{ext}}=\SI{-15}{mT}\) is applied along the \(z\)-axis to keep the YIG film in a saturated state without destabilizing the DWs in the stripe. We use standard values of the saturation magnetization and exchange stiffness for YIG (\(M_\mathrm{s} = 139~\mathrm{kA\,m^{-1}}\), \(A_\mathrm{ex} = 4~\mathrm{pJ\,m^{-1}}\)) and Py (\(M_\mathrm{s} = 800~\mathrm{kA\,m^{-1}}\), \(A_\mathrm{ex} = 13~\mathrm{pJ\,m^{-1}}\)). The Gilbert damping constants are \(\alpha = 1 \times 10^{-5}\) in YIG and \(\alpha = 0.008\) in Py.

\begin{figure*}[t]
\centering
\includegraphics[width=\linewidth]{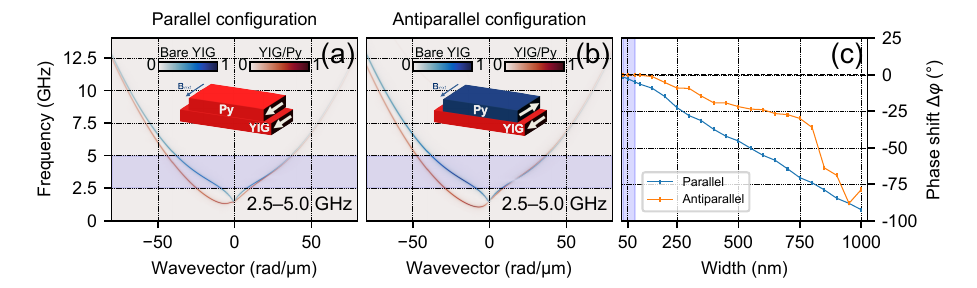}
\caption{Selecting the operating frequency regime. (a,b) Dispersion of a bare YIG film (blue) and YIG/Py bilayer (red) for two orientations of the magnetization in the Py layer: (a) parallel and (b) antiparallel to the magnetization of YIG.
(c) The SW phase shift \(\Delta \varphi\) induced by a uniformly magnetized Py stripe is shown as a function of the stripe's width \(w\) for the parallel and antiparallel orientations of the stripe's magnetization relative to that of YIG, the bias magnetic field \(B_{\mathrm{ext}}=\SI{-15}{mT}\) and the frequency \(f=2.8\)~GHz. The transparent blue regions indicate the working frequency and width ranges.
}
  \label{Fig:dispersions}
\end{figure*}

Numerical simulations are carried out using MuMax3~\cite{Vansteenkiste2014}. The system is discretized using \( 5 \times 5 \times 5~\mathrm{nm^3}\) cells. To avoid effects of the demagnetization at the edges of the simulated volume, periodic boundary conditions (PBC) were applied along the \(x\) (4 repetitions) and \(z\)-axis (256 repetitions). 
SWs are excited on the left side of the YIG film using a localized, microwave-frequency magnetic field designed to emit a monochromatic Damon-Eshbach (DE) SW propagating toward the racetrack region, along the $+x$ axis. Detailed information on the SW antenna implementation is provided in the Supplementary Materials, Sec.~SI.B. 
To suppress SW reflections from the boundaries and approximate an open propagation channel of width smaller than the separation between the DWs, we use absorbing boundary conditions (ABCs). The latter are implemented using YIG regions in which the Gilbert damping is gradually ramped up toward the edges.
A damping increase in the transverse directions was used to confine SWs to a $220\,\mathrm{nm}$ wide channel. This is narrower than the distance between the pinned DWs, so that the SWs mainly propagate underneath a single DW-bounded domain, where the stray field is most uniform.
The transverse ABCs absorb all energy leaving the channel sideways — including waves diffracted at the DWs and scattered by the neighboring domains — regardless of their propagation direction.
As a result, the wave reaching the extraction region is dominated by the front propagating along $x$ with a nearly flat transverse phase, which lets us isolate the phase imprinted by a single domain and avoid a strong transverse phase gradient across the analyzed wavefront.
The complementary case, in which diffraction at an aperture and the DWs is not suppressed, is examined in Sec.~\ref{Sec:RealisticSImulations}.
Further details of the damping profile are provided in the Supplementary Material, Sec.~SI.A.


\section{Results\label{Sec:Results}}
\begin{figure*}[t]
\centering
\includegraphics[width=\linewidth]{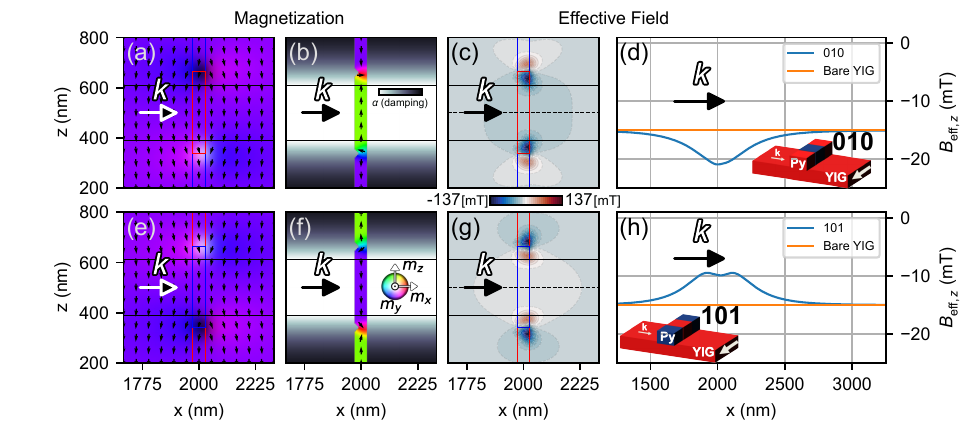}
\caption{Magnetization and effective-field landscapes underlying the DW-induced phase shifts. 
(a,e) Equilibrium magnetization texture in the YIG layer for (a) \(\mathrm{010}\) and (e) \(\mathrm{101}\) racetrack configurations shown in panels (b) and (f), respectively, for the racetrack width \(w = 50~\mathrm{nm}\) and magnetic field \(B_{\mathrm{ext}}=\SI{-15}{mT}\) along the $z$-axis. The displayed magnetization maps correspond to the top surfaces of the YIG film and the racetrack. (c,g) Corresponding maps of the effective field component \(B_{\mathrm{eff},z}\) in the YIG film with the stray field of the racetrack. In (b) and (f), the Gilbert damping parameter increased from the YIG nominal value in the middle to 0.5 at the film edges, as shown by the grey colour code. (d,h) Corresponding one-dimensional cuts of the effective field \(B_{\mathrm{eff},z}\) along the central axis of the SW channel (dashed lines in panels (c) and (g)) with and without the racetrack.}
\label{Fig:mag_field}
\end{figure*}
\begin{figure*}[t]
\centering
\includegraphics[width=\linewidth]{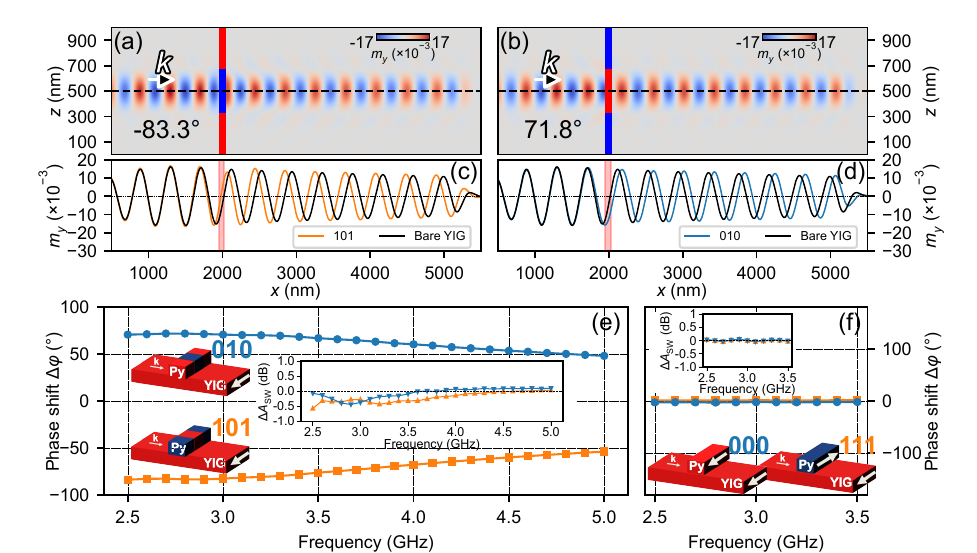}
\caption{Dynamic profiles of SWs propagating in the YIG channel for two DW configurations of the racetrack: (a,c) \(\mathrm{101}\) and (b,d) \(\mathrm{010}\). 
(a,b) Spatial maps of the dynamic magnetization component \(m_y\) showing SW propagation at frequency 3.0~GHz across the DW region (vertical bar), with the dashed lines indicating the position of the profiles shown in (c) and (d). (c,d) Corresponding one-dimensional \(m_y(x)\) profiles of SWs from the YIG film with (orange and blue lines) and without (black lines) the racetrack. (e) Frequency dependence of the accumulated phase shift of SWs \(\Delta \varphi\) for the \(\mathrm{101}\) and \(\mathrm{010}\) configurations of the racetrack. Inset shows the frequency dependence of the transmitted SW intensity attenuation.
(f) Phase shift for uniformly magnetized racetracks in \(\mathrm{000}\) and \(\mathrm{111}\) configurations. The racetrack width is \(w = 50~\mathrm{nm}\), and the magnetic field \(B_{\mathrm{ext}}=\SI{-15}{mT}\) is applied along the $z$-axis.
}
\label{Fig:phase}
\end{figure*}


We begin by identifying a frequency regime in which the racetrack without the DWs, i.e., only a uniformly magnetized stripe, does not produce a substantial SW phase shift in the YIG film. Figures~\ref{Fig:dispersions}(a) and~\ref{Fig:dispersions}(b) show the SW dispersion relations calculated for a YIG/Py bilayer with the same 5~nm separation between the layers and infinite extent along the \(+x\) axis. Panels (a) and (b) correspond to the Py magnetization oriented parallel and antiparallel to the magnetization of the YIG layer, respectively. On the positive-\(k\) branch, in the selected operating frequency range \(2.5\)--\(5~\mathrm{GHz}\), the bilayer dispersion is close to that of plain YIG because the mutual alignment of the bias field and the SW propagation direction concentrates the SW intensity near the bottom surface of the YIG film, further decreasing the effect of dynamic dipolar coupling within the bilayer region~\cite{DAMON1961308, Qin2021_nature}. This indicates that the perturbation of SWs propagating in the \(+x\) direction by the uniformly magnetized Py layer is weak. 

This agrees with the results of simulations when the width of the Py stripe is finite. Figure~\ref{Fig:dispersions}(c) shows the phase shift \(\Delta\varphi\) acquired by SWs at a frequency \(f=2.8\)~GHz as they propagate under a uniformly magnetized Py stripe as a function of the stripe width \(w\). (For all geometries and magnetic configurations considered throughout the paper, the phase shifts are evaluated relative to a corresponding reference simulation without the racetrack for the same excitation frequency and extraction region.) For homogeneously magnetized Py stripes with widths smaller than 80 nm, the observed SW phase shift is tiny (less than $5^\circ$), both in the parallel and antiparallel configurations. The shift increases with the stripe width, reaching approximately \(-92^\circ\) and \(-78^\circ\) for the parallel and antiparallel magnetization states, respectively, for the width of \(1000~\mathrm{nm}\). For the antiparallel configuration, we observe a strong deviation of the \(\Delta\varphi(w)\) dependence from a straight line. This deviation occurs due to the presence of a magnonic Fabry--P\'{e}rot resonance (MFPR) in the system and its interaction with the propagating SW modes~\cite{Lutsenko2025}. The frequency of the MFPR strongly depends on the width of the stripe; therefore, for wider racetracks, MFPR-related effects may require separate consideration. In this work, we focus on the phenomena that are not related to the MFPR, and restrict the racetrack width to \(w \leq 80~\mathrm{nm}\) while keeping its thickness fixed at \(20~\mathrm{nm}\). This choice ensures that the frequency of the lowest MFPR mode stays above the selected operating range of 2.5–5 GHz, and the phase shift produced by a uniformly magnetized racetrack remains below approximately \(5^\circ\), which is much smaller than that due to the DWs in the racetrack, as we discuss below. 


We now turn to the racetrack containing two DWs, schematically shown in Fig.~\ref{Fig:structure}(b). The DWs separate domains with magnetization aligned either parallel or antiparallel to that in YIG and denoted as $0$ and $1$, respectively. The two multi-domain configurations of the racetrack can then be encoded as $010$ and $101$, while \(\mathrm{000}\) and \(\mathrm{111}\) describe its configurations with a uniform magnetization. The two DWs have volume magnetic charges of opposite sign, which generate a stray magnetic field in the underlying YIG layer. As shown in Fig.~\ref{Fig:mag_field}, this stray field modifies locally the internal magnetic field and the total effective magnetic field in the YIG film. The $010$ / $101$ configurations result in a local increase / decrease of the modulus of effective magnetic field in YIG; consequently, SWs should experience a slower / faster phase variation as they propagate under the racetrack. 

To evaluate the effect of the DWs on the SW propagation, we extract the SW phase shift and attenuation as a function of frequency for four magnetization states of the racetrack: \(\mathrm{000}\), \(\mathrm{111}\), \(\mathrm{101}\), and \(\mathrm{010}\).
The attenuation is quantified as the intensity ratio, calculated as the squared ratio of the root-mean-square dynamic magnetization amplitude in the transmitted-wave region with and without the racetrack (again, using the same spatial and temporal averaging windows in both cases). In the plots, this intensity ratio is expressed in decibels.

Let us consider the racetrack with a width of 50 nm. For its homogeneous states of the racetrack, \(\mathrm{000}\) and \(\mathrm{111}\) (Fig.~\ref{Fig:phase}(f)), the phase shift remains close to zero 
over the entire 2.5--3.5\,GHz window, which is consistent with the negligible modification of the bilayer dispersion relation for the forward propagation direction (Fig.~\ref{Fig:dispersions}). In contrast, the two configurations with DWs yield much larger phase shifts, ranging from $70.70^{\circ}$ to $48.10^{\circ}$ for the \(\mathrm{010}\) configuration and from $-83.7^{\circ}$ to $-53.77^{\circ}$ for the \(\mathrm{101}\) configuration in the 2.5--5.0\,GHz frequency range. The signs of the phase shifts are opposite, as expected. For both configurations, the attenuation remains close to 0~dB (i.e. does not exceed -1~dB) in the considered frequency range change (see the inset in Fig.~\ref{Fig:phase}e). 

The observed phase shift dependence on the domain configuration in the racetrack can be qualitatively understood by analyzing the equilibrium magnetization and effective-field distributions in the YIG film under the racetrack (Fig.~\ref{Fig:mag_field}). The magnetic charges of the DWs generate positive and negative lobes of the effective field component $B_{\mathrm{eff},z}$ underneath the racetrack. 
This field change effectively increases or decreases the SW wavelength locally, depending on the magnetic configuration: \(\mathrm{010}\) or \(\mathrm{101}\), corresponding to increasing and decreasing modulus of transverse stray field component, respectively (see Fig.~\ref{Fig:mag_field}). 
Although the racetrack width is only \(w=50~\mathrm{nm}\), the DW-induced perturbation of \(B_{\mathrm{eff},z}\) extends along the SW propagation direction over a much larger distance, as shown in Fig.~\ref{Fig:mag_field}(d,h). 


A SW phase shift can also arise from the stray magnetic field generated by the magnetic charges at the ends of a Py stripe of finite length (nanoelement). The nanoelement can also be switched between two micromagnetic states: \(\mathrm{-0-}\) (magnetization parallel to YIG) and \(\mathrm{-1-}\) (antiparallel), as shown in the Supplementary material, Fig.~S2. As for the DW-based case, the shift is negative (positive) in the parallel (antiparallel) state. However, the total magnetic charge at one end of the nanoelement is half that in one DW in a racetrack with the same cross-section. Consequently, the effects observed for a nanoelement are about half weaker than those for a pair of DWs in the racetrack. Indeed, for the nanoelement of 320 nm length, the maximum accumulated phase shift is around $40^\circ$ at 2.5 GHz, whereas it is around $80^\circ$ for the racetrack. The nanoelement can also be mechanically actuated, as was done for permanent micromagnets and magnetic flux concentrators, used to tune the effective field and so to control the SW phase in a CoFeB magnonic waveguide~\cite{Cocconcelli2025}. Yet, the racetrack-based approach  proposed here is more flexible and scalable. Indeed, the DWs can be displaced by current pulses ~\cite{Parkin2008,Bahri2019} between pinning sites, which can form a rather dense mesh.


\begin{figure*}[t]
\centering
\includegraphics[width=\linewidth]{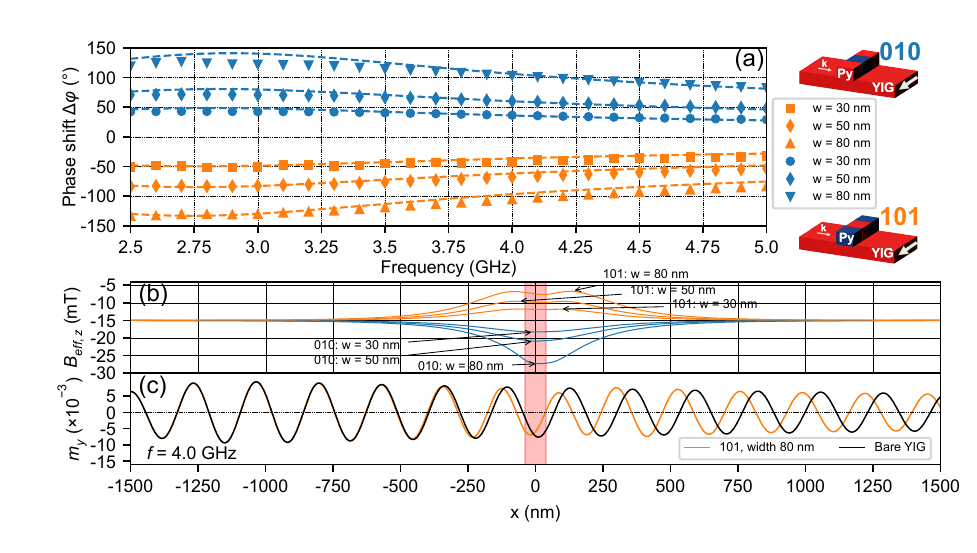}
\caption{Frequency dependence of the SW phase shift \(\Delta\varphi\) for racetracks of 30, 50, and 80~nm width and \(\mathrm{101}\) and \(\mathrm{010}\) configurations. 
(a) Phase shift \(\Delta\varphi(f)\) obtained from full MuMax3 simulations (markers) compared with predictions of the WKB-based semi-analytical model (dashed lines). 
(b) Corresponding effective-field profiles \(B_{\mathrm{eff},z}(x)\) for different racetrack widths and both configurations obtained from simulations. 
(c) Spatial profiles of the dynamic magnetization component \(m_y(x)\) for a racetrack width of \(80~\mathrm{nm}\) in the 101 configuration  at \(f = 4.0~\mathrm{GHz}\), illustrating the accumulated phase shift relative to the reference signal (black).}
\label{Fig:theory}
\end{figure*}

\section{Semi-analytical Model Of Domain-Wall-Induced Phase Shift\label{Sec:AnalyticalModel}}
To gain a deeper insight in the observed phase shift, we use a semi-analytical model based on the Wentzel–Kramers–Brillouin (WKB) approximation. This semiclassical method is typically applied to wave propagation in slowly varying potentials, including SWs in graded magnonic landscapes \cite{Schlomann1964,Demokritov2004,Fripp2021}. In our case, the effective magnetic field is strongly nonuniform in the YIG film (Fig.~\ref{Fig:mag_field}). So, we average the simulated field over the YIG thickness ($y$-direction) and then consider the central slice of the obtained distribution along the $z$-direction as one-dimensional profile $B_{\mathrm{eff},z}(x)$. 
Then, the total phase shift accumulated by the propagating SW is defined as
\begin{align}
\Delta \varphi(f) =  \int_{-\infty}^{\infty} \left[ k(f, B_{\mathrm{eff},z}(x)) - k_0(f, B_\mathrm{ext}) \right] \, dx,
\label{eq:Delta_phi}
\end{align}
where \( k(f, B_{\mathrm{eff},z}(x)) \) is the position-dependent SW wave number corresponding to the field profile \( B_{\mathrm{eff},z}(x)\) and \( k_0 \) is the unperturbed wave number at the uniform external field value \( B_\mathrm{ext} = -15 \) mT and frequency \( f \). Both $k$ and $k_0$ are obtained from Kalinikos-Slavin analytical dispersion relation~\cite{Kalinikos_1986} for the fundamental mode in the DE geometry.  The details are given in the Supplementary Material, Sec.~SIII, including the method of obtaining \( B_{\mathrm{eff},z}(x)\) and the field profiles used are shown in Fig.~\ref{Fig:theory}(b). 



The model’s predictions are compared with simulations across different racetrack widths in Fig.~\ref{Fig:theory}(a). The analytical model (dashed lines) agrees well with the micromagnetic simulation results (symbols), capturing the large phase shifts over a broad frequency range for 30~nm, 50~nm and 80~nm widths. This confirms that the observed phase shift of the propagating SWs is due to a locally modified internal magnetic field under the racetrack as well as associated changes in the dispersion relation. 

The increase of the phase-shift magnitude with the racetrack width can be understood from the effective-field profiles in Fig.~\ref{Fig:theory}(b).
The racetrack width affects the magnitude of the stray-field variation more than its spatial extent: wider racetracks produce a deeper / higher $B_{\mathrm{eff},z}(x)$ lobe but a similar longitudinal extent.
It is this enhancement of the $B_{\mathrm{eff},z}(x)$ magnitude that is primarily responsible for the larger accumulated phase shift in wider racetracks, a trend that the model reproduces well.
For context, in the investigated range ($2.5-5.0~\mathrm{GHz}$) the unperturbed SW wavelength decreases from approximately 760 to 170~nm, so the racetrack widths stay below the SW wavelength across the band, while the extent of the DW-induced perturbation in Fig.~\ref{Fig:theory}(b) is comparable to, or larger than, the wavelength, as illustrated by the SW profile at $f=4.0$~GHz in Fig.~\ref{Fig:theory}(c).
A noticeable discrepancy between the model and the simulations emerges only for the 80~nm-wide racetrack, reaching up to approximately
$18^\circ$ for the 010 configuration around $f \approx 3.0~\mathrm{GHz}$,
and about $9^\circ$ for the 101 configuration around
$f \approx 4.3~\mathrm{GHz}$. Two factors contribute to this.
First, for the widest racetrack the stray field is strongest and most manifestly two-dimensional, so the single central 1D slice used in the model no longer fully represents the perturbation acting on the propagating SW.
Second, at the lower end of the band the wavelength ($\sim760$ nm at 2.5 GHz) becomes comparable to the longitudinal extent of the perturbation, so only a few wavelengths span the varying-field region and the slowly-varying-field assumption underlying the WKB approximation is least accurate.

\begin{figure*}[t]
\centering
\includegraphics[width=\linewidth]{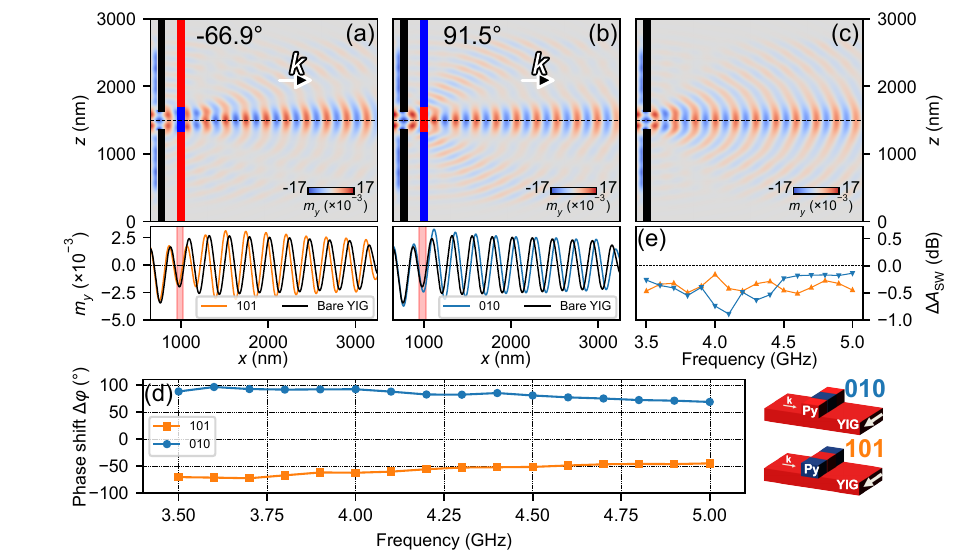}
\caption{SW phase shift in a single aperture--racetrack geometry. Spatial distributions of the dynamic magnetization \(m_y\) at 4 GHz for (a) \(\mathrm{101}\) and (b) \(\mathrm{010}\), and (c) bare YIG (without the racetrack) configurations. The dashed horizontal line marks the cross-section used for phase extraction. Bottom row: \(m_y(x)\) profiles corresponding to the distributions in panels (a) and (b) in comparison to that in panel (c). The profiles show opposite phase shifts relative to the reference, equal to approximately \(-66.9^\circ\) for 101 and \(+91.5^\circ\) for 010. (d) Frequency dependence of the phase shift \(\Delta\varphi\); insets show the corresponding racetrack configurations. The bias magnetic field \(B_{\mathrm{ext}}=-15~\mathrm{mT}\) is applied along the \(z\)-axis and (e) frequency dependence of the transmitted SW intensity normalized to the reference. 
}
\label{Fig:phase_diffraction_grating}
\end{figure*}
\section{Spin-Waves phase shifting in a single aperture--racetrack geometry\label{Sec:RealisticSImulations}}

Let us now verify that the phase-shifting effect persists in geometries where SWs are not artificially confined to a channel formed by ABCs. 
Instead, the SWs are excited as a plane wave and pass through a non-magnetic ($M_\mathrm{s}=0$) screen of 50~nm length along the $x$ axis, with a $300$~nm opening along the $z$ axis, placed $250$~nm before a $50$~nm wide racetrack in the $\mathrm{010}$ or $\mathrm{101}$ configuration (Fig.~\ref{Fig:phase_diffraction_grating}(a),(b)).
In contrast to the channel geometry of the previous sections, the low-damping active region here is much wider ($\sim\!2.2~\mu\mathrm{m}$), so that the diffracting beam can also reach the DWs and the neighboring domains rather than propagating only under the central domain.
The aperture width is 300~nm, while the domain of reversed magnetization in the racetrack is  \(\mathrm{320}\)~nm long. After passing through the aperture, an initially plane SW starts diverging (Fig.~\ref{Fig:phase_diffraction_grating}(c)), due to diffraction.  
Different parts of the SW wave front then experience different phase shifts as they pass under the racetrack. The simulated SW maps and the corresponding line profiles (Figs.~\ref{Fig:phase_diffraction_grating}(a-c)) show that the phase-shifting effect is preserved. The frequency dependence of the SW phase shift for the aperture simulations is presented in Fig.~\ref{Fig:phase_diffraction_grating}(d). As before, the two magnetic states result in phase shifts of opposite signs, and their strengths remain comparable to those obtained in previous simulations (Fig.~\ref{Fig:phase}(a-d)). For instance, at 4.0~GHz, the extracted phase shift is approximately \(-66.9^\circ\) for the \(\mathrm{101}\) configuration and \(+91.5^\circ\) for the \(\mathrm{010}\) configuration. The transmitted intensity remains close to the reference level. 

The SW amplitude maps from Fig. \ref{Fig:phase_diffraction_grating}(a-c) show that the change of the racetrack DW configuration affects the beam spreading. This is confirmed by the reciprocal-space maps of SWs at 4 and 5~GHz (obtained by 2D Fourier transforms of the SW maps from the YIG area behind the racetrack to the momentum space \cite{Whitehead2018}) presented in Fig.~S3 in the Supplementary Material, show the intensity distribution along the isofrequency contour  to be dependent on the magnetic configuration. The \(\mathrm{101}\) configuration results in smaller defocusing, while the \(\mathrm{010}\) configuration produces only a slight increase in the angular spread, as compared to the reference case. 

The positive values of the transmitted SW intensity ratio in dB present in Fig.~\ref{Fig:phase}(e) are not related to real SW amplification. It comes from the different beam shape after passing the racetrack region. The \(\mathrm{101}\) configuration slightly concentrates the beam, which reduces lateral spreading and losses into the absorbing boundaries, so the measured intensity can be higher than in the reference case. In contrast, the \(\mathrm{010}\) configuration slightly broadens the beam, so this effect is smaller. This boundary-related increase is not observed in the single-aperture geometry shown in Fig.~\ref{Fig:phase_diffraction_grating}.

\section{Conclusions}
We have proposed and numerically validated a SW phase shifter based on a narrow magnetic DW racetrack integrated with a YIG film. Within the $2.5$--$5.0$~GHz operating band, multi-domain configurations of opposite polarity lead to robust phase shifts of opposite signs, while the uniform (i.e. DW-free) configurations of the same racetrack exhibit negligible phase shifts. Effective-magnetic field maps confirm that the phase shifts originate from the stray-field imprint produced by the DWs in the YIG film under the racetrack, which modifies the SW wavelength and so the phase that they accumulate upon transmission through this region. The results of our micromagnetic simulations agree with those obtained using a WKB-based model, which maps the simulated three-dimensional field distribution onto an effective one-dimensional profile along the propagation direction. The model captures the main trends of the simulated phase response and provides a transparent link between the DW stray field and the resulting phase accumulation. Our results demonstrate that a racetrack with sub-wavelength width can operate as a programmable SW phase shifter. This suggests a route toward scalable functional elements required for SW-based conventional and emerging unconventional computing paradigms where information is encoded and processed in the phase of propagating SWs.

\begin{acknowledgments}
The research leading to these results was funded by the European Union’s Horizon Europe research and innovation program under GA 101070347 MANNGA. Views and opinions expressed are however those of the authors only and do not necessarily reflect those of the European Union or the European Health and Digital Executive Agency (HADEA). Neither the European Union nor the granting authority can be held responsible for them. This work was also supported by the National Science Centre, Poland, Projects: OPUS No.~UMO-2023/49/B/ST3/02920, SONATA, UMO-2024/55/D/ST3/02079,
PRELUDIUM No.~UMO-2024/53/N/ST3/03244. The contribution of the Norwegian Financial Mechanism 2014-2021, Project POLS No. UMO-2020/37/K/ST3/02450 is also acknowledged.
\end{acknowledgments}

\section*{Data Availability Statement}
The data supporting the findings of this study are openly available on Zenodo under DOI:
\href{https://doi.org/10.5281/zenodo.20704687}{10.5281/zenodo.20704687}.

\section*{AUTHOR DECLARATIONS}

\subsection*{Conflict of Interest}
The authors have no conflicts to disclose.

\subsection*{Author Contributions}
\textbf{U.M.}: Conceptualization; Funding acquisition; Methodology (lead); Software; Investigation (micromagnetic simulations); Formal analysis; Visualization; Data curation; Writing -- original draft.
\textbf{O.T.}: Funding acquisition; Methodology; Validation.
\textbf{P.G.}: Funding acquisition; Writing -- review \& editing  (supporting).
\textbf{A.L.}: Conceptualization; Writing -- review \& editing  (supporting).
\textbf{S.v.D.}: Conceptualization; Funding acquisition; Writing -- review \& editing (supporting). 
\textbf{V.V.K.}: Conceptualization; Funding acquisition; Writing -- review \& editing (lead). 
\textbf{M.K.}: Supervision; Conceptualization; Methodology (supporting); Writing -- review \& editing (lead).

\bibliography{refs}

\newpage
\clearpage
\onecolumngrid

\largeSection{SUPPLEMENTARY MATERIAL}

The supplementary material provides additional details on the simulation setup and data analysis. It includes the absorbing-boundary-condition profiles, antenna implementation, finite-length Permalloy stripe reference simulations, details of the WKB model, and reciprocal-space analysis of spin waves in the single-aperture racetrack geometry.

\vspace{1em}

\suppnumbering

\section{Geometry and the simulation method}
\label{Sec:S_SimGeometryParams}

\subsection{Absorbing boundary condition}
\label{Sec:S_ABC}
\begin{figure*}[t]
\centering
\includegraphics[width=\linewidth]{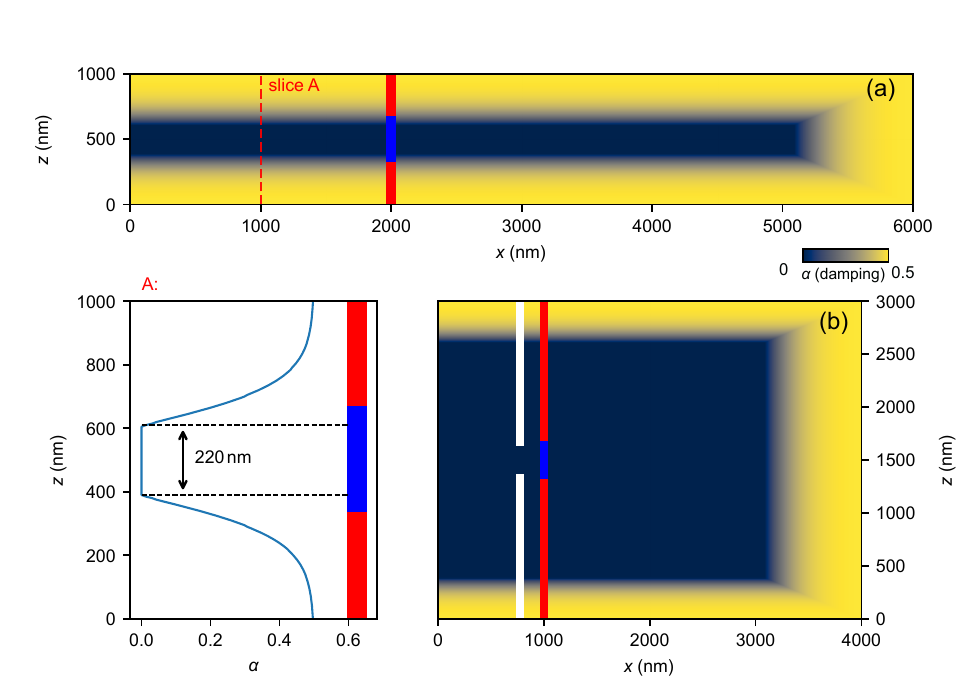}
    \caption{
    Spatial distribution of the Gilbert damping parameter $\alpha$ used to implement the absorbing boundary conditions. 
    The color scale shows the normalized damping profile, from the active-region value $\alpha = 1\times10^{-5}$ to the absorbing-frame value $\alpha = 0.5$. 
    (a) ABC profile used for the straight racetrack geometry. The dashed red line marks slice A, whose transverse damping profile is shown on the left. The dashed horizontal lines in the slice profile indicate the $220\,\mathrm{nm}$-wide active spin-wave channel. 
    (b) ABC profile used for the single-aperture (white bars) diffraction geometry. 
    The dark regions correspond to the active propagation areas, while the bright regions indicate the absorbing damping frames. 
    The red vertical line with the blue segment marks the racetrack/domain position.
    }
\label{Fig:ABC}
\end{figure*}

In MuMax3 simulations~[S1], we implement absorbing boundary conditions (ABC) in the form of damping frames introduced at the lateral edges of the simulation window to avoid reflection of spin waves (SWs): along the $y$-axis, and additionally along the $x$ direction to form a SW propagation channel of width comparable to DW separation. The YIG layer has dimensions
\(6000\times1000\times50\,\mathrm{nm}^3\). As shown in Fig.~\ref{Fig:ABC}(a), the damping is increased near the transverse edges of the simulation window and at the right end of the waveguide. The graded damping profile is defined as
\begin{align}
\alpha_i=\alpha_0+\alpha_{\mathrm{add}}
\tanh\!\left[3\frac{i}{N_{\mathrm{grad}}}\right],
\end{align}
where \(\alpha_0=10^{-5}\), \(\alpha_{\mathrm{add}}=0.5\), and \(N_{\mathrm{grad}}=230\). Here, \(i\) is an integer index running from
\(0\) to \(N_{\mathrm{grad}}\). The ramp is implemented using spatial steps of \(\Delta x=0.8~\mathrm{nm}\) along the propagation direction and \(\Delta z=0.34~\mathrm{nm}\) along the transverse direction.

The transverse damping frames leave a central low-damping SW propagation channel of approximately \(220\,\mathrm{nm}\) width. In the standard racetrack geometry, Fig.~\ref{Fig:ABC}(a), the two inner notches are located at \(z=500\pm160~\mathrm{nm}\), giving a notch-to-notch separation of \(320~\mathrm{nm}\). Thus, the active SW channel is fully contained within the DW-bounded racetrack region. 
For the single-diffraction geometry (Fig.~\ref{Fig:ABC}(b))
the aperture is formed by a non-magnetic ($M_\mathrm{s}=0$) region of 300~nm opening, and the surrounding low-damping active region is much wider ($\sim2.2~\mu\mathrm{m}$), bounded by the same ABC frames as before, so that the SW beam can diffract after the aperture before reaching the absorbing frames.

\subsection{Antenna design\label{Sec:Antenna}}
The excitation of SWs contains \(x\)- and \(y\)-components of the microwave magnetic field with a relative phase shift \(\Delta\phi=-\pi/2\), forming a rotating field that efficiently couples to the magnetization dynamics. The spatial profile consists of a sinusoidal modulation along the propagation and vertical in-plane directions with a Gaussian envelope:
\begin{align}
B_{x,y}(x,z,t) &= B_1 \sin[\omega(t-t_0)]\,\sin\!\big(kx+\phi_{x,y}\big) \\
&\times \exp\!\left[-\frac{(x-x_s)^2}{2\sigma_x^2}
-\frac{(z-z_s)^2}{2\sigma_z^2}\right],
\end{align}
where \(\phi_x=\Delta\phi\) and \(\phi_y=0\). Here \(k=2\pi/\lambda\), where \(\lambda\) is obtained from the Kalinikos--Slavin dispersion relation~[S2]. The parameters \(\sigma_x\) and \(\sigma_z\) define the spatial confinement of the antenna along the propagation and transverse directions, respectively, and are defined as
\[
\sigma_x = \lambda,
\qquad
\sigma_z = \frac{\SI{240}{\; nm}}{2\sqrt{2\ln 2}}.
\]

\section{Finite-length Permalloy stripe as a phase shifter of propagating spin waves\label{Sec:finite_stripe}}
Here we consider a single uniformly magnetized Permalloy nanostripe of finite length. This geometry is kept identical to the racetrack geometry used for the DW-based phase shifter, except that the racetrack parts outside the central domain are removed. The nanostripe has dimensions \(50\times320\times20~\mathrm{nm}^3\), where \(50~\mathrm{nm}\) is its width along the SW propagation direction, \(320~\mathrm{nm}\) is its length along the transverse direction, and \(20~\mathrm{nm}\) is its thickness. It is placed above a \(50~\mathrm{nm}\)-thick YIG layer, with a  separation of \(5~\mathrm{nm}\). The external magnetic field is set to \(B_{ext}=-15~\mathrm{mT}\) and applied along \(z\)-direction. 
Two magnetic states are analyzed: magnetization parallel and antiparallel to the magnetization in the YIG layer.

\begin{figure*}[t]
\centering
\includegraphics[width=\linewidth]{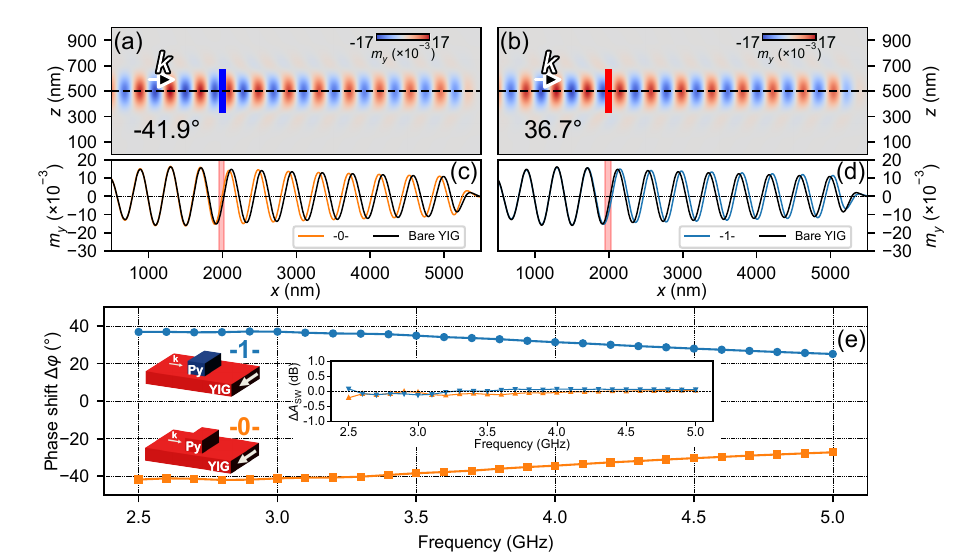}
\caption{Dynamic profiles of SWs propagating in the YIG channel under the Py-stripe of 320~nm length in two magnetization configurations: (a, c) parallel (left column) and (b, d) antiparallel (right column) to the magnetization of YIG film at 3~GHz. 
(a,b) Spatial maps of the dynamic magnetization component \(m_y\) showing SW propagation below the Py nanostripe (vertical bar), with the horizontal dashed line indicating the position of the extracted profiles in (c) and (d). 
(c,d) Corresponding one-dimensional \(m_y(x)\) profiles compared with the reference signal (SW propagating in a bare YIG film, without a nanostripe). 
(e) Frequency dependence of the accumulated phase shift \(\Delta \varphi\) for the parallel and antiparallel configurations of the magnetization in the nanostripe. Inset shows the frequency dependence of the transmitted SW intensity attenuation.
}
\label{Fig:phase_lim}
\end{figure*}

The results are summarized in Fig.~\ref{Fig:phase_lim}. In both configurations, the finite stripe produces phase modulation with only a weak change in the transmitted intensity in comparison with effect of the racetrack with DWs (Fig.~\ref{Fig:phase_lim}(e)--inset). The sign of the phase shift is the same as in the DW racetrack case: the parallel state yields a negative phase shift, whereas the antiparallel state yields a positive one. However, the magnitude of the phase shift is about half smaller. At \(f=3.0\) GHz, the uniformly magnetized finite stripe gives phase shifts of approximately \(-41.9^\circ\) and \(+36.7^\circ\) (Fig.~\ref{Fig:phase_lim}(a,b)), whereas the racetrack containing a finite domain segment between two domain walls gives about \(-83.3^\circ\) and \(+71.8^\circ\) for the 101 and 010 states, respectively. Thus, for the same longitudinal size of the stripe and the domain limited by two DWs, the DW-based racetrack produces a phase response that is roughly two times larger than that of the uniformly magnetized finite stripe. 

\section{WKB model of the domain-wall-induced phase shift\label{Sec:WKB}}

To support the micromagnetic simulations, we use a semi-analytical model based on the Wentzel--Kramers--Brillouin (WKB) approximation. The purpose of the model is to relate the static stray-field landscape generated by the DWs in the Py racetrack to the accumulated phase of a SW propagating in the underlying YIG film at a fixed frequency.

The one-dimensional effective field profile used in the WKB calculation is constructed from the static demagnetizing field obtained in MuMax3. Since the bias field and the equilibrium magnetization are oriented along the \(z\)-axis, we use the \(z\)-component of the racetrack demagnetizing field and add the uniform external field. The field profile is extracted at the central slice \(z=z_{\mathrm c}\) of the waveguide and averaged over the selected YIG thickness along the \(y\)-direction. Thus, 
\begin{equation} B_{\mathrm{eff},z}(x) = B_{\mathrm{ext}} + \left\langle B_{\mathrm{demag},z}(x,y,z_{\mathrm c}) \right\rangle_{y}, \label{eq:Beff} \end{equation} 
where \(\langle\cdots\rangle_y\) denotes averaging over the selected thickness of the YIG layer,  \(B_\mathrm{ext}\) is the uniform bias magnetic field, \(\mathbf{B}_{\mathrm{demag}}\) is the static demagnetizing field produced by the racetrack.

For a given frequency \(f\), the local wave vector \(k(x)\) is obtained from the Kalinikos--Slavin dispersion relation for the fundamental thickness mode. In the form used in the calculations,
\begin{equation}
f=\frac{1}{2\pi}\sqrt{\omega_{0}(k,x)\left[\omega_{0}(k,x)+\omega_{M}F(k,\phi_k(x))\right]}
\label{eq:KS_main}
\end{equation}
with
\begin{equation}
\omega_{0}(k,x)=\gamma\left[\left|B_{\mathrm{eff},z}(x)\right|+\frac{2A_{\mathrm{ex}}}{M_{\mathrm{s}}}k^{2}\right],
\qquad
\omega_{M}=\gamma\mu_{0}M_{s},
\label{eq:omega0}
\end{equation}
where \(\gamma\) is the gyromagnetic ratio, \(A_{\mathrm{ex}}\) is the exchange stiffness, \(M_{s}\) is the saturation magnetization, and \(\mu_{0}\) is the free-space permeability.

The dipolar term is written as
\begin{equation}
F(k,\phi_k)=P(kt)+\sin^{2}\theta
\left[
1-P(kt)\left(1+\cos^{2}\phi_k\right)
+\frac{\omega_{M}}{4\omega_{0}}\sin^{2}\phi_k\left(1-e^{-2kt}\right)
\right],
\label{eq:Fterm}
\end{equation}
where \(t\) is the YIG thickness, and \(\theta\) and \(\phi_k\) define the orientation of the equilibrium magnetization with respect to the film normal and the SW propagation direction, respectively. The thickness-dependent function \(P(kt)\) is
\begin{equation}
P(kt)=1-\frac{1-e^{-kt}}{kt}.
\label{eq:Pkt}
\end{equation}
In the Damon--Eshbach geometry considered here, the equilibrium magnetization lies in the film plane and is perpendicular to the SW propagation direction, so that \(\theta=\pi/2\) and \(\phi_k=\pi/2\).

For each frequency, the local wave vector \(k(x)\) is found numerically by solving Eq.~(\ref{eq:KS_main}) for the field profile \(B_{\mathrm{eff},z}(x)\). The reference wave vector \(k_{0}\) is obtained in the same way for the unperturbed field \(B_{0}\). The WKB phase shift is then calculated as
\begin{align}
\Delta \varphi(f) =  \int_{-x_1}^{x_2} \left[ k(f, B_{\mathrm{eff},z}(x)) - k_0(f, B_\mathrm{ext}) \right] \, dx,
\label{eq:sup_delta_phi}
\end{align}
where \(x_{1}\) and \(x_{2}\) define the numerical integration window covering the racetrack perturbation region. A local increase in the modulus of \(B_{\mathrm{eff},z}(x)\) decreases the local wave vector and leads to a positive accumulated phase shift (see Fig.~\ref{Fig:theory}(b) blue lines), whereas a local decrease of modulus of \(B_{\mathrm{eff},z}(x)\) increases the local wave vector and produces a negative phase shift  (see Fig.~\ref{Fig:theory}(b) orange lines).


\begin{figure*}[b]
\centering
\includegraphics[width=\linewidth]{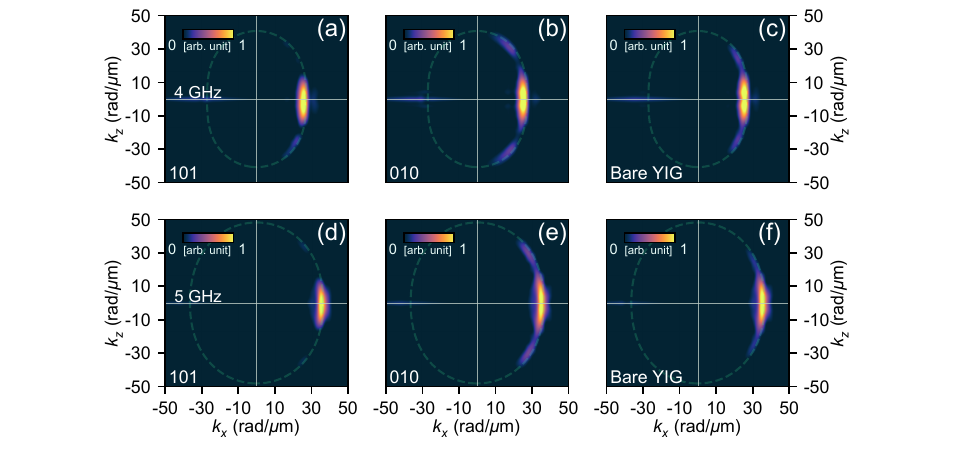}
\caption{Reciprocal-space distributions of the SW at 4~GHz (a--c) and 5~GHz (d--f) for the \(\mathrm{101}\), \(\mathrm{010}\) and the bare YIG. The bias magnetic field \(B_{\mathrm{ext}}=\SI{-15}{mT}\) is applied along the $z$-axis.}
\label{Fig:isofreq}
\end{figure*}

If the local frequency were to fall below the FMR, the wave vector would become complex and the WKB solution would additionally contain an evanescent attenuation factor proportional to
\(\exp[-\int \Im k(x;f)\,dx]\). However, this regime is outside the operating window analyzed in the main text.

This semi-analytical approach contains no adjustable dynamic parameters beyond the field profile extracted from the micromagnetic ground state and the material parameters of YIG. It therefore provides a direct link between the DWs stray field and the extracted phase shift. At the same time, the model has clear limitations: it neglects backscattering, mode conversion, and the full three-dimensional character of the perturbation. These approximations are responsible for the residual discrepancies observed for the widest racetrack, especially for the 101 configuration, where the stray field extends further in space and the perturbation is less accurately represented by a one-dimensional reduction.

\section{Reciprocal-space analysis of spin waves in a single aperture--racetrack geometry\label{Sec:k_space}}
To analyze the angular distribution of the transmitted SW beam, we performed a two-dimensional Fourier transform of the dynamic magnetization maps recorded in the YIG region immediately downstream of the racetrack, i.e., on the transmitted side of the waveguide. Only data from the steady-state regime.
The reciprocal-space distributions in Fig.~\ref{Fig:isofreq} show the Damon--Eshbach isofrequency contour. The 101 configuration produces a more concentrated distribution, while the 010 configuration gives a slightly broader angular spread. Thus, the racetrack state affects not only the accumulated SW phase but also the beam redistribution.

\section*{References}

\noindent\hangindent=1.6em\hangafter=1
[S1] A. Vansteenkiste, J. Leliaert, M. Dvornik, M. Helsen, F. Garcia-Sanchez, and B. Van Waeyenberge,
The design and verification of mumax3,
AIP Advances \textbf{4}, 107133 (2014).

\noindent\hangindent=1.6em\hangafter=1
[S2] B. A. Kalinikos and A. N. Slavin,
Theory of dipole-exchange spin wave spectrum for ferromagnetic films with mixed exchange boundary conditions,
Journal of Physics C: Solid State Physics \textbf{19}, 7013 (1986).

\end{document}